\documentclass[aps,prl,twocolumn,superscriptaddress,showpacs,floatfix]{revtex4}
\usepackage{graphicx,epsfig,amsmath}

\newcommand{\bohrm}{\mbox{$\mu_{B}$}}
\newcommand{\kboltz}{\mbox{$k_{B}$}}

\newcommand{\AlxGaAs}[2]{\mbox{$\text{Al}_{#1}\text{Ga}_{#2}\text{As}$}}
\newcommand{\Tone}{\mbox{$\text{T}_{\text{1}}$}}
\newcommand{\Ttwo}{\mbox{$\text{T}_{\text{2}}$}}
\newcommand{\ZeemanE}{\mbox{$|g|\bohrm B$}}

\newcommand{\startsubfig}[2]{Figure~\ref{fig:#1}(#2)}
\newcommand{\subfig}[2]{Fig.~\ref{fig:#1}(#2)}
\newcommand{\allfig}[1]{Fig.~\ref{fig:#1}}
\newcommand{\ket}[1]{\mbox{$\left|#1\right\rangle$}}

\begin{document}


\title{Electrical control of spin relaxation in a quantum dot}

\author{S. Amasha}
	\email{samasha@mit.edu}
	\affiliation{Department of Physics, Massachusetts Institute of Technology, Cambridge, Massachusetts 02139}

\author{K. MacLean}
	\affiliation{Department of Physics, Massachusetts Institute of Technology, Cambridge, Massachusetts 02139}

\author{Iuliana P. Radu}
	\affiliation{Department of Physics, Massachusetts Institute of Technology, Cambridge, Massachusetts 02139}

\author{D. M. Zumb\"{u}hl }
	\affiliation{Department of Physics and Astronomy, University of Basel, Klingelbergstrasse 82, CH-4056 Basel, Switzerland}

\author{M. A. Kastner} 
	\affiliation{Department of Physics, Massachusetts Institute of Technology, Cambridge, Massachusetts 02139}
	
\author{M. P. Hanson} 
	\affiliation{Materials Department, University of California, Santa Barbara 93106-5050}	

\author{A. C. Gossard} 
	\affiliation{Materials Department, University of California, Santa Barbara 93106-5050}


\begin{abstract}
	We demonstrate electrical control of the spin relaxation time \Tone~between Zeeman split spin states of a single electron in a lateral quantum dot. We find that relaxation is mediated by the spin-orbit interaction, and by manipulating the orbital states of the dot using gate voltages we vary the relaxation rate $W \equiv \Tone^{-1}$ by over an order of magnitude. The dependence of $W$ on orbital confinement agrees with theoretical predictions and from these data we extract the spin-orbit length. We also measure the dependence of $W$ on magnetic field and demonstrate that spin-orbit mediated coupling to phonons is the dominant relaxation mechanism down to $1$ T, where $\Tone$ exceeds $1$ s.
\end{abstract}


\maketitle


	Achieving macroscopic control of quantum states has become an important part of developing systems for applications in quantum computing and spintronics \cite{Loss1998:QDotQuanComp, Awschalom2007:spintronics}. Control of the spin states of individual electrons confined in quantum dots is of particular interest \cite{Hanson2006:SpinsinQD}. In a magnetic field $B$ the spin states of the electron are split by the Zeeman energy $\Delta= \ZeemanE$, providing a two level quantum system that can be used as a qubit for quantum computing \cite{Loss1998:QDotQuanComp} or as the basis of spin memory \cite{Kroutvar2004:OpticalMemoryQD}. Recent experiments have demonstrated the ability to manipulate \cite{Petta2005:CoherentManipulation, Koppens2006:ESR} and read-out \cite{Elzerman2004:SingleShotReadOut, Hanson2005:SpinDepTunnelRates} the electron's spin. An important remaining challenge is to better understand and control the interactions between the electron's spin and its solid-state environment.	
	
		The two most important of these are the hyperfine and spin-orbit interactions. The hyperfine interaction causes decoherence of the spin states by coupling the electron's spin to Ga and As nuclear spins \cite{Petta2005:CoherentManipulation, Khaetskii2002:ElecSpinDecoh, Ono2004:NucSpin, Koppens2005:STMixing, Johnson2005:TSrelax}; however, because the effective nuclear magnetic field changes slowly, coherent behavior is still observed \cite{Petta2005:CoherentManipulation}. Recently methods have been suggested for suppressing the hyperfine-induced decoherence \cite{Klauser2006:SpinNarrowing, Giedke2006:QMeas, Stepanenko2006:OptEnhance}.

	The spin-orbit interaction (SOI) causes spin relaxation by mixing the orbital and spin states, thus providing a mechanism for coupling the spin to electric fluctuations in the environment of the dot \cite{Khaetskii2001:ZeemanSF, Golovach2004:PhononInducedDecay, Falko2005:Anisotropy, Stano2006:PhononRelax, SanJose2006:GeoSpin, Marquardt2005:SpinRelaxNyquist, Borhani2006:QPCSpinDecay}, specifically to piezoelectric phonons \cite{Khaetskii2001:ZeemanSF, Golovach2004:PhononInducedDecay, Falko2005:Anisotropy, Stano2006:PhononRelax, SanJose2006:GeoSpin}. This coupling induces spin relaxation and brings the probabilities of being in the excited and ground spin states to thermal equilibrium; at temperatures $T \ll \Delta/\kboltz$ an electron can relax from the excited to the ground spin state by emission of a piezoelectric phonon. The timescale for energy relaxation is \Tone, and since relaxation necessarily destroys any coherent spin state, it sets a limit $\Ttwo<2\Tone$ \cite{Golovach2004:PhononInducedDecay}. By measuring \Tone~\cite{Elzerman2004:SingleShotReadOut, Fujisawa2002:AllowForbid, Hanson2003:SpinRelaxation, Amasha2006:LowFieldT1} and varying the energy between the spin states, it has been shown that this mechanism accounts for relaxation between two-electron triplet and singlet states \cite{Hanson2005:SpinDepTunnelRates, Sasaki2005:PumpProbe, Meunier2007:STRelax} in lateral GaAs dots, as well as for spin relaxation between one electron Zeeman sublevels in a layer of self-assembled Ga(In)As quantum dots \cite{Kroutvar2004:OpticalMemoryQD}.  
	
	In this Letter we demonstrate \textit{in situ} electrical control of the spin relaxation rate of a single electron in a lateral quantum dot by using gate voltages to manipulate the mixing of the spin and orbital states. This allows us to vary the spin relaxation rate $W\equiv \Tone^{-1}$ by over an order of magnitude at fixed $\Delta$. We find that $W$ depends only on the confinement of the electron wavefunction in the direction along the applied in-plane magnetic field as expected for the SOI in GaAs, and that the dependence of $W$ on the energy scale for confinement is that predicted by theory \cite{Khaetskii2001:ZeemanSF, Golovach2004:PhononInducedDecay}. From these data we extract the spin-orbit length, which describes the strength of the SOI. We also measure $W$ as a function of field down to $1$ T, where we find that \Tone~is longer than $1$ s, and demonstrate that spin-orbit mediated phonon-induced spin decay is the dominant relaxation mechanism in single-electron lateral dots down to low magnetic fields.

\begin{figure}[]
\begin{center}
\includegraphics[width=8.0cm, keepaspectratio=true]{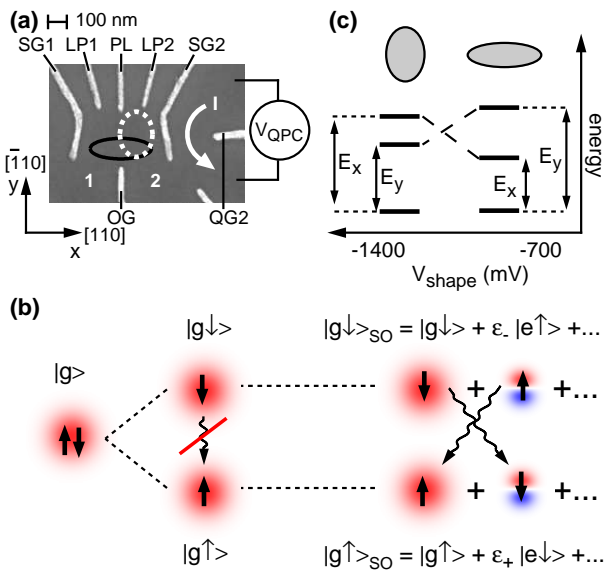}
\end{center}

	\caption{(color online) (a) Electron micrograph of the gate geometry. Negative voltages are applied to the labeled gates while the unlabeled gate and the ohmic leads (which are numbered) are kept at ground. The black solid (white dotted) ellipse illustrates the expected dot shape for less (more) negative $V_{shape}$. The magnetic field is parallel to the $y$-axis and all voltage pulses are applied to gate LP2. (b) At $B= 0$ and with no SOI, the spin-$\uparrow$ and spin-$\downarrow$ states of the ground orbital state $\ket{g}$ are degenerate. Applying a magnetic field splits the spin states but phonon coupling between \ket{g\uparrow} and \ket{g\downarrow} is prohibited. The SOI acts as a perturbation and mixes the orbital and spin states: the perturbed spin states $\ket{g\uparrow}_{SO}$ and $\ket{g\downarrow}_{SO}$ contain excited orbital states (\ket{e}) of the opposite spin so the perturbed states can be coupled by phonons. The SOI involves a momentum operator and requires a change in parity for coupling of different orbital states. (c) Dot energy spectrum as gate voltages are varied to change the shape of the dot. The value of $V_{shape}$ is the voltage on SG1 for a given set of gate voltages.
}
	\label{fig:dotstates}
\end{figure}

	The dot used in this work is fabricated from an \AlxGaAs{0.3}{0.7}/GaAs heterostructure grown by molecular beam epitaxy. The two-dimensional electron gas formed at the material interface $110$ nm below the surface has a density of $2.2\times10^{11}$ $\text{cm}^{-2}$ and a mobility of $6.4\times10^{5}$ $\text{cm}^{2}$/Vs \cite{Granger2005:TwoStageKondo}. To form a single-electron quantum dot we apply negative voltages to metallic gates patterned on the surface (\subfig{dotstates}{a}). Adjacent to the dot is a quantum point contact (QPC) charge sensor \cite{Field1993:NoninvasiveProbe}: when an electron tunnels onto the dot the negative charge increases the resistance of the QPC, which we measure by sourcing a current $I$ and measuring the change in voltage $\delta V_{QPC}$. We adjust the gate voltages to make the tunneling rate between the dot and lead $2$ slower than the bandwidth of the charge sensing circuit (the tunneling rate to lead $1$ is kept negligibly small). This allows us to observe electron tunneling events in real time \cite{Elzerman2004:SingleShotReadOut, Lu2003:RealTimeDet, Gustavsson2006:CountStat, MacLean2007:WKB}. Measurements are made in a dilution refrigerator with an electron temperature of $120$ mK (unless noted otherwise) and the magnetic field is applied parallel to the AlGaAs/GaAs interface so it does not affect the dot's orbital states.

	Spin relaxation in single-electron quantum dots involves excited orbital states (\subfig{dotstates}{b}). The dominant mechanism for exchanging energy with the environment is for the electron to interact with electrical fluctuations from piezoelectric phonons \cite{Khaetskii2001:ZeemanSF, Golovach2004:PhononInducedDecay, Falko2005:Anisotropy, Stano2006:PhononRelax, SanJose2006:GeoSpin}. However, while phonons can couple different orbital states of the dot, they cannot couple different spin states. Coupling between spin states is made possible by the SOI which mixes the Zeeman split ground orbital state with excited orbital states of the opposite spin \cite{Khaetskii2001:ZeemanSF}. This allows phonons to induce spin relaxation as illustrated in \subfig{dotstates}{b}. By changing the energy of the excited orbital states, we can control the amount of SOI induced mixing, and thus control the spin relaxation rate.  

	Using the gate voltages, we manipulate the dot shape and hence its orbital states. We model the electrostatic potential of the dot with an anisotropic harmonic oscillator potential $U(x,y) = \frac{1}{2} m^{*} \omega_{x}^{2} x^{2} + \frac{1}{2} m^{*} \omega_{y}^{2} y^{2}$. When the voltages on all dot gates are roughly equal, one expects from the gate geometry that the dot is less confined along the $x$-axis than along $y$ (black solid ellipse in \subfig{dotstates}{a}). Consequently, the lowest lying excited state is at energy $E_{x}= \hbar\omega_{x}$ above the ground state, while the next higher excited state has $E_{y}= \hbar\omega_{y}$ (assuming $E_{y} < 2E_{x}$). We introduce the parameter $V_{shape}$ to describe how we change the dot shape: more negative $V_{shape}$ corresponds to a more negative voltage on SG1, which pushes the dot towards SG2 and increases confinement along $x$. More negative $V_{shape}$ also corresponds to a less negative voltage on LP1, PL, and LP2, which reduces confinement along $y$ (white dotted ellipse in \subfig{dotstates}{a}) as well as compensating for the SG1 voltage change to keep the ground state energy constant. The numerical value of $V_{shape}$ given in \subfig{dotstates}{c} is the voltage on SG1 for the set of gate voltages. These geometric considerations lead us to expect that $E_{x}$ should increase and $E_{y}$ should decrease as $V_{shape}$ is made more negative (\subfig{dotstates}{c}).

\begin{figure}[]
\begin{center}
\includegraphics[width=8.0cm, keepaspectratio=true]{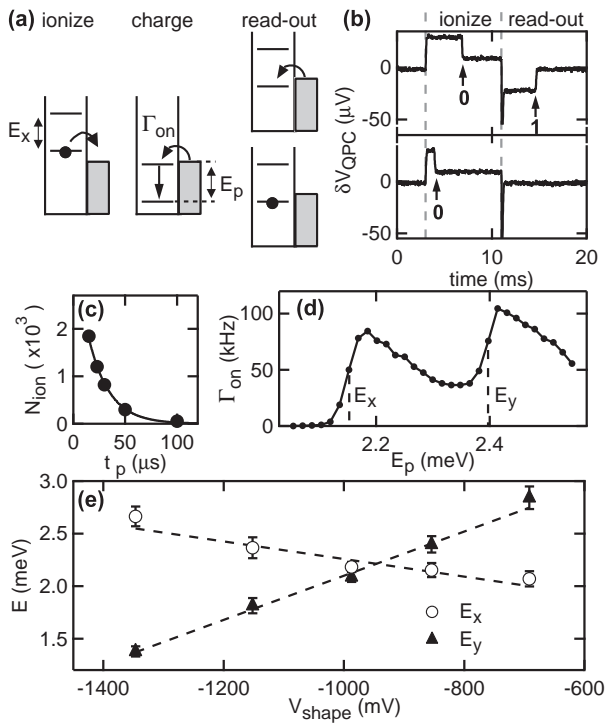}
\end{center}

	\caption{(a) Three step pulse sequence for measuring the energy of the excited orbital states. (b) Examples of real-time data. The direct capacitive coupling to the pulsed gate causes the QPC to respond to the pulse sequence; electron tunneling events are evident on top of this response. The 0's denote when an electron tunnels off the dot, while 1's denote when an electron tunnels on. The charging pulse ($t_{p} = 50 \mu$s for this example) appears as a sharp spike between the ionization and read-out periods. (c) The number of events $N_{ion}$ for which the dot is empty after the charging pulse (top panel in \subfig{tspec}{b}) as a function of $t_{p}$. The solid line is a fit to an exponential to determine the rate $\Gamma_{on}$ at which electrons tunnel onto the dot. (d) $\Gamma_{on}$ vs $E_{p}$ for $V_{shape}= -850$ mV. The two sharp rises mark the energies when an excited state crosses the Fermi energy. (e) Energies of the excited orbital states of the dot as a function of $V_{shape}$. The dashed lines are linear fits to the data.
}
	\label{fig:tspec}
\end{figure}

	At each $V_{shape}$ we measure the energy of the excited orbital states using a three step pulse sequence (\subfig{tspec}{a}) with B= 0. After ionizing the dot, we apply a pulse $V_{p}$ to bring the ground orbital state an energy $E_{p} = e\alpha_{LP2} V_{p}$ below the Fermi energy of the lead, where $e \alpha_{LP2}$ is a conversion factor we calibrate for each $V_{shape}$ \cite{footnote:alphas, Gustavsson2006:superPoisson}. We find that $e \alpha_{LP2}$ increases as $V_{shape}$ is made more negative as we expect from the geometric considerations discussed in \allfig{dotstates}. We apply the pulse for time $t_{p}$ that is short ($15~\mu\mbox{s}<t_{p}<400~\mu\mbox{s}$) compared to the average tunneling time into the ground state ($\approx 10$ ms near the Fermi energy), so the probability for tunneling into the ground orbital state is small. However, for sufficiently large $E_{p}$ one or more excited orbital states will be below the Fermi energy. These states are more strongly coupled to the leads than the ground state \cite{Gustavsson2006:CountStat, MacLean2007:WKB}, and an electron can tunnel onto the dot with rate $\Gamma_{on}$. Once on, the electron quickly decays to the ground state \cite{Fujisawa2002:AllowForbid, Climente2006:ElecRelax}. 

	Finally, in the read-out state we position the ground state just below the Fermi energy. If the dot is still ionized then an electron tunnels onto the dot (top right in \subfig{tspec}{a}) and we observe this with our real-time charge detection system (top panel of \subfig{tspec}{b}). We count the number of times $N_{ion}$ this occurs and find $N_{ion}$ decreases exponentially with $t_{p}$ (\subfig{tspec}{c}). The rate of decay gives $\Gamma_{on}$ and \subfig{tspec}{d} shows $\Gamma_{on}$ as a function of $E_{p}$. The two large increases correspond to the energies at which the excited orbital states cross the Fermi energy. As the excited states are pulled below the Fermi energy with increasing $E_{p}$, $\Gamma_{on}$ decreases because the energy of the excited state is decreased relative to the height of the tunnel barrier \cite{MacLean2007:WKB}. \startsubfig{tspec}{e} shows the energies at several values of $V_{shape}$, and shows one state increasing and one state decreasing in energy. This behavior is what we expect from the geometric considerations: as the confinement along $x$ increases and along $y$ decreases with more negative $V_{shape}$, the energy $E_{x}$ of the $x$-excited state increases, while the energy $E_{y}$ of the $y$-excited state decreases, allowing us to identify the $x$ and $y$ states as indicated in \subfig{tspec}{e}. This orientation of the dot orbital states is also consistent with our spin relaxation measurements discussed next. 

\begin{figure}[]
\begin{center}
\includegraphics[width=8.0cm, keepaspectratio=true]{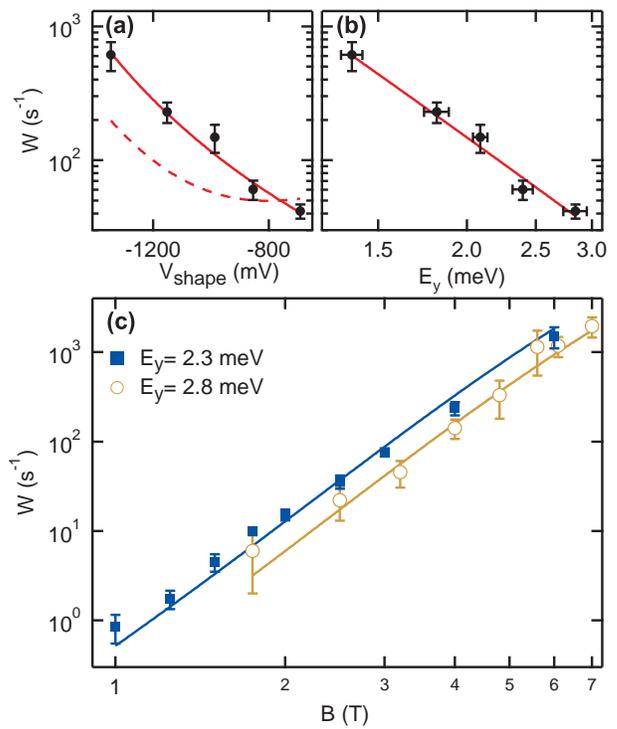}
\end{center}

	\caption{(color online) (a) $W$ as a function of $V_{shape}$ at $B= 3$ T. The solid and dashed lines are fits discussed in the text. (b) The same relaxation rate data as in \subfig{T1data}{a}, plotted as a function of $E_{y}$. The solid line is a fit to find the spin-orbit length as discussed in the text. (c) Spin relaxation rate as a function of magnetic field for two different sets of gate voltages. Solid lines are fits discussed in the text.
}
	\label{fig:T1data}
\end{figure}

	For each $V_{shape}$, we measure $W\equiv\Tone^{-1}$ at $B= 3$ T. To do this, we first ionize the dot and then pulse both Zeeman-split levels below the Fermi level for a time $t_{w}$. During this time electrons can tunnel onto the dot and then relax from the excited to the ground spin state. By measuring the decay of the probability of being in the excited spin state as a function of $t_{w}$, we obtain $W$ \cite{Amasha2006:LowFieldT1}. The results are shown in \subfig{T1data}{a} demonstrating we can electrically control $W$ by over an order of magnitude. We have verified that the Zeeman splitting does not vary with $V_{shape}$, confirming the observed variation is caused by changes in the orbital states rather than coupling to phonons of different energy. 

	The energy of the excited orbital states affect $W$ because the higher the energy of the excited states, the weaker the SOI coupling to the ground state, and the slower the relaxation rate. If we model $W$ assuming the potential $U(x,y)$ given above, an in-plane $B$, a SOI that is linear in the electron's momentum, and a phonon wavelength much greater than the dot size then $W= A_{x}E_{x}^{-4} + A_{y}E_{y}^{-4}$. Here $A_{x}$ and $A_{y}$ describe the contribution of each orbital to spin relaxation and $W\propto E^{-4}$ because of Van Vleck cancellation \cite{Khaetskii2001:ZeemanSF}. We fit the data in \subfig{T1data}{a} to this equation by approximating $E_{x}(V_{shape})$ and $E_{y}(V_{shape})$ by the dashed lines shown in \subfig{tspec}{e}. The solid line in \subfig{T1data}{a} shows the result of the fit and we find that $A_{x}/A_{y} < 14\%$, implying that only the y-excited orbital state is contributing to spin relaxation. The dashed line shows a fit where we require $A_{x} = A_{y}$; clearly the data is inconsistent with both the $x$ and $y$-excited states contributing equally. 

	We can understand why the $y$-excited state dominates spin relaxation from the spin-orbit Hamiltonian. In the coordinate system from \subfig{dotstates}{a} it takes the form $H_{SO} = (\beta-\alpha)p_{y}\sigma_{x} + (\beta+\alpha)p_{x}\sigma_{y}$ where $\alpha$ and $\beta$ are the Rashba and Dresselhaus spin-orbit parameters, respectively \cite{Golovach2004:PhononInducedDecay}. $B$ is applied along the $y$-axis so only the first term in $H_{SO}$, which is proportional to  $\sigma_{x}$, can couple different spin states as in \subfig{dotstates}{b}. Since this term is proportional to $p_{y}$, a change in parity along the $y$-axis is also required. The $x$-excited state does not satisfy this requirement, so the $p_{y}\sigma_{x}$ term couples the Zeeman split ground orbital state to $y$-excited states of opposite spin. A consequence is that for $V_{shape} > -1000$ mV, it is the higher energy excited state that determines $W$, an unusual situation.

	For a comparison to theory, \subfig{T1data}{b} shows $W$ as a function of $E_{y}$; here the directly measured values of $E_{y}$ are used. In the limit where the phonon wavelength is much larger than the size of the dot, $W \approx A B^{5} E_{y}^{-4} \lambda_{SO}^{-2}$ where $A = 33~\mbox{s}^{-1}\mbox{meV}^{4}\mu\mbox{m}^{2}/\mbox{T}^{5}$ depends on previously measured values of $|g|$ and phonon parameters in GaAs, and $\lambda_{SO} = \hbar/m^{*}|\beta - \alpha|$. We fit the data in \subfig{T1data}{b} to a theoretical prediction by Golovach et al. \cite{Golovach2004:PhononInducedDecay} that includes the effects of the phonon wavelength being comparable to the size of the dot and obtain $\lambda_{SO} = 1.7 \pm 0.2~\mu$m, which is consistent with previous measurements in dots \cite{Zumbuhl2002:SOCoupling}. We note that taking $E_{x}$ instead of $E_{y}$, i.e. a different dot orientation, to explain the spin relaxation would be inconsistent with the data ($W$ would increase with increasing $E_{x}$), independently confirming the dot orientation as deduced from geometric considerations.

	Spin relaxation also depends sensitively on the magnetic field \cite{Kroutvar2004:OpticalMemoryQD} as shown in \subfig{T1data}{c} for two different sets of gate voltages. At $1$ T, $W$ is less than $1~\mbox{s}^{-1}$, corresponding to a very long $\Tone > 1$ s. These data demonstrate electrical control of $W$ over a range of magnetic fields. The solid lines through the data are fits to the theory of Golovach et al. \cite{Golovach2004:PhononInducedDecay} using our value for $\lambda_{SO}$. The fits give values of $E_{y}$ consistent with what we expect for these gate voltage configurations and the rather large $E_{y}$ chosen allows for very long spin lifetimes. At low magnetic fields other spin relaxation mechanisms may become important, such as SOI mediated coupling to electrical fluctuations from the ohmic leads \cite{SanJose2006:GeoSpin}, surface gates \cite{Marquardt2005:SpinRelaxNyquist}, and the QPC \cite{Borhani2006:QPCSpinDecay} or hyperfine coupling to phonons \cite{Erlingsson2002:HyperfineRelax}. However, the agreement between our data and theory down to a field of $1$ T demonstrates that spin-orbit mediated coupling to piezoelectric phonons is the dominant spin relaxation mechanism in single electron lateral quantum dots down to low magnetic fields, corresponding to very long times.
		
	In summary, we have demonstrated electrical control of the spin relaxation rate of a single electron in a lateral quantum dot by manipulating the orbital states \textit{in situ} using gate voltages. The measured dependence of $W$ on orbital confinement and magnetic field is in excellent agreement with theory \cite{Khaetskii2001:ZeemanSF, Golovach2004:PhononInducedDecay}, demonstrating that spin-orbit mediated coupling to phonons is the dominant spin relaxation mechanism. 

	We are grateful to V. N. Golovach, D. Loss, and L. Levitov for discussions, to V. N. Golovach for providing his code to perform calculations, and to I. J. Gelfand and T. Mentzel for experimental help. This work was supported by the US Army Research Office under W911NF-05-1-0062, by the National Science Foundation under DMR-0353209, and in part by the NSEC Program of the National Science Foundation under PHY-0117795.


\begin{thebibliography}{37}
\bibitem{footnote:alphas}
\bibinfo{note}{We calibrate $e \alpha_{LP2}$ by measuring the width of the Fermi
  function of ohmic lead 2 at $300$ mK.}
\end{thebibliography}

\end{document}